\documentclass[11pt,twoside]{article}

\usepackage{asp2006}
\usepackage{graphics,graphicx}

\begin{document}

\title{The Case of the 300 kpc Long X-ray Jet in PKS~1127-145 at z=1.18}   

\author{Aneta Siemiginowska, D.E.Harris, Thomas L. Aldcroft}   

\affil{Harvard-Smithonian Center for Astrophysics 60 Garden St., 
Cambridge, MA 02138}    

\author{{\L}ukasz Stawarz,C.C.Cheung\altaffilmark{1}}

\affil{Kavli Institute for Particle Astrophysics and Cosmology, 
Stanford University, Stanford CA 94305}

\author{Marek Sikora}
\affil{Copernicus Astronomical Center, Warsaw, Poland}

\author{Jill Bechtold} 
\affil{Steward Observatory, University of Arizona, Tucson, AZ}

\altaffiltext{1}{Jansky Postdoctoral Fellow; National Radio Astronomy Observatory}

\begin{abstract} 

The complex X-ray morphology of the 300 kpc long X-ray jet in
PKS~1127-145 (z=1.18 quasar) is clearly discerned in a $\sim$100 ksec {\it
Chandra} observation. The jet X-ray surface brightness gradually
decreases by an order of magnitude going out from the core.  The X-ray
spectrum of the inner jet is relatively flat with
$\alpha_X=0.66^+_-0.15$ and steep in the outer jet with
$\alpha_X=1.0^+_-0.2$. The X-ray and radio jet intensity profiles are
strikingly different, with the radio emission peaking strongly at the
two outer knots while the X-ray emission is strongest in the inner jet
region.  We discuss the constraints implied by these data on the X-ray
emission models and conclude that ``one-zone'' models fail and that at
least a two-component model is needed to explain the jet's broadband
emission. We propose that the X-ray emission originates in the jet
proper while the bulk of the radio emission comes from a surrounding
jet sheath. We also consider intermittent jet activity as a possible
cause of the observed jet morphology.

\end{abstract}




\section{Introduction}

The origin of the X-ray emission in quasar jets is puzzling
\citep[see][for review]{harris2006} because a straight
extrapolation of the synchrotron radio-to-optical continuum into the
X-rays underpredicts the luminosity of powerful {\it Chandra} large
scale jets. Thus, the same single power-law population of electrons
cannot produce the radio, optical, and X-ray emission in the framework
of a homogeneous one-emission zone approximation \citep[e.g.][]{sambruna2004}
The synchrotron self-Compton (SSC) process cannot easily
explain the data because it does not produce enough X-rays at the
equipartition fields and therefore requires large departures from the
minimum-power condition \citep{chartas2000,harris2002,kataoka2005}.
The alternative scenario involves
inverse-Compton scattering of the Cosmic Microwave Background (IC/CMB)
photons implying large jet bulk Lorentz factors ($\Gamma$) at hundreds
of kpc from the active nuclei \citep{tavecchio2000,celotti2001,schwartz2002}.
In this paper we discuss the constraints on
the X-ray emission models given by the {\it Chandra} and VLA
observations of the PKS~1127-145 jet.

\section{Jet Properties}

\begin{figure}[t]
\begin{center}
\includegraphics[width=7.8cm]{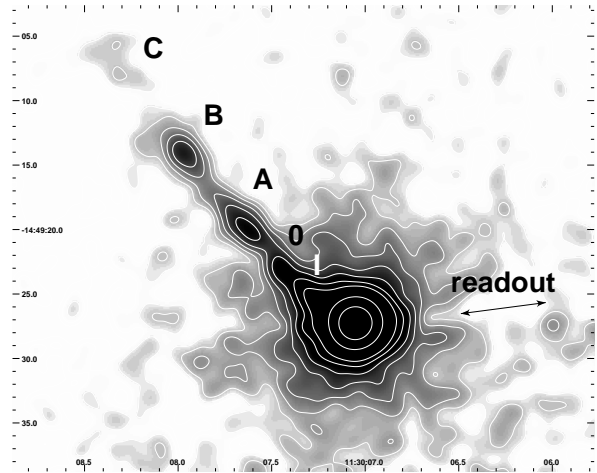}
\caption{{\it Chandra} ACIS-S exposure corrected image of PKS1127--145
(E=0.3-7~keV) and smoothed with the Gaussian kernel
(1$\sigma$=0.615~arcsec).   Contours are factor of two in log scale starting at
10$^{-10}$ photons~cm$^{-2}$~s$^{-1}$~pixel$^2$. The two final contours
within the core are at 10$^-6$ and 10$^-7$
photons~cm$^{-2}$~s$^{-1}$~pixel$^2$ (1 pixel=0.123'').}
\end{center}
\label{xrays}
\end{figure}

One of the longest X-ray jets known (see
\texttt{http://hea-www.harvard.edu/XJET/}) is associated with the
redshift z=1.18 radio-loud quasar PKS~1127-145 (Fig.~1). This jet
poses several questions for X-ray emission models
\citep[see][]{siemi2002,siemi2007}. The jet morphology is
complex with several knots seen in the X-ray image (Fig.~1) with the
furthest knot C being the weakest. The prominent knots A and B are
connected by continuous X-ray emission that stops beyond knot B at
$\sim 22$~arcsec from the core. The VLA radio maps show low brightness
emission along the jet (Fig.~2). The one-sided jet shows an X-ray
surface brightness which declines with the distance from the core,
while the radio brightness increases (Fig.~3[left]).  The X-ray to
radio intensity ratio decreases along the jet. Fig.3[right] shows the
changes to spectral index along the jet. The radio spectrum is steep
and X-ray spectrum is flat in the inner jet regions.  There are only
upper limits to the knots optical brightness, but they are too high
for constraining the emission models.  Thus only radio and X-ray
observations can be used to study emission processes in this jet.  We
discuss the constraints on the models below.

\begin{figure}
\begin{center}
\includegraphics[width=7.6cm]{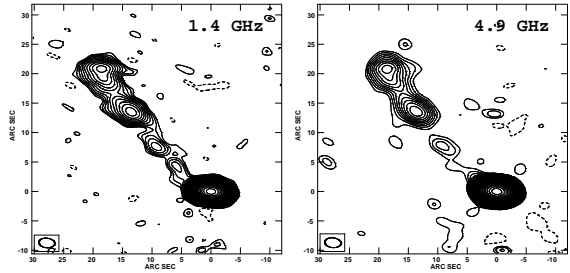}
\caption{VLA 1.4 and 4.9 GHz images of PKS~1127--145. The lowest contour levels
plotted are  0.75 mJy beam$^{-1}$ (1.4 GHz) and
0.39 mJy beam$^{-1}$ (4.9 GHz) and increase by 
factors of $\sqrt{2}$ up to peaks of 5.46 and 4.14 Jy/beam, 
respectively. The (uniform weighted) beam sizes are:
2.71\arcsec$\times$1.62\arcsec\ at PA=84.4$^{\circ}$,
and 2.77\arcsec$\times$1.66\arcsec\ at PA=84.0$^{\circ}$, respectively.}
\end{center}
\label{radio}
\end{figure}

\begin{figure}[t]
\includegraphics[width=7cm]{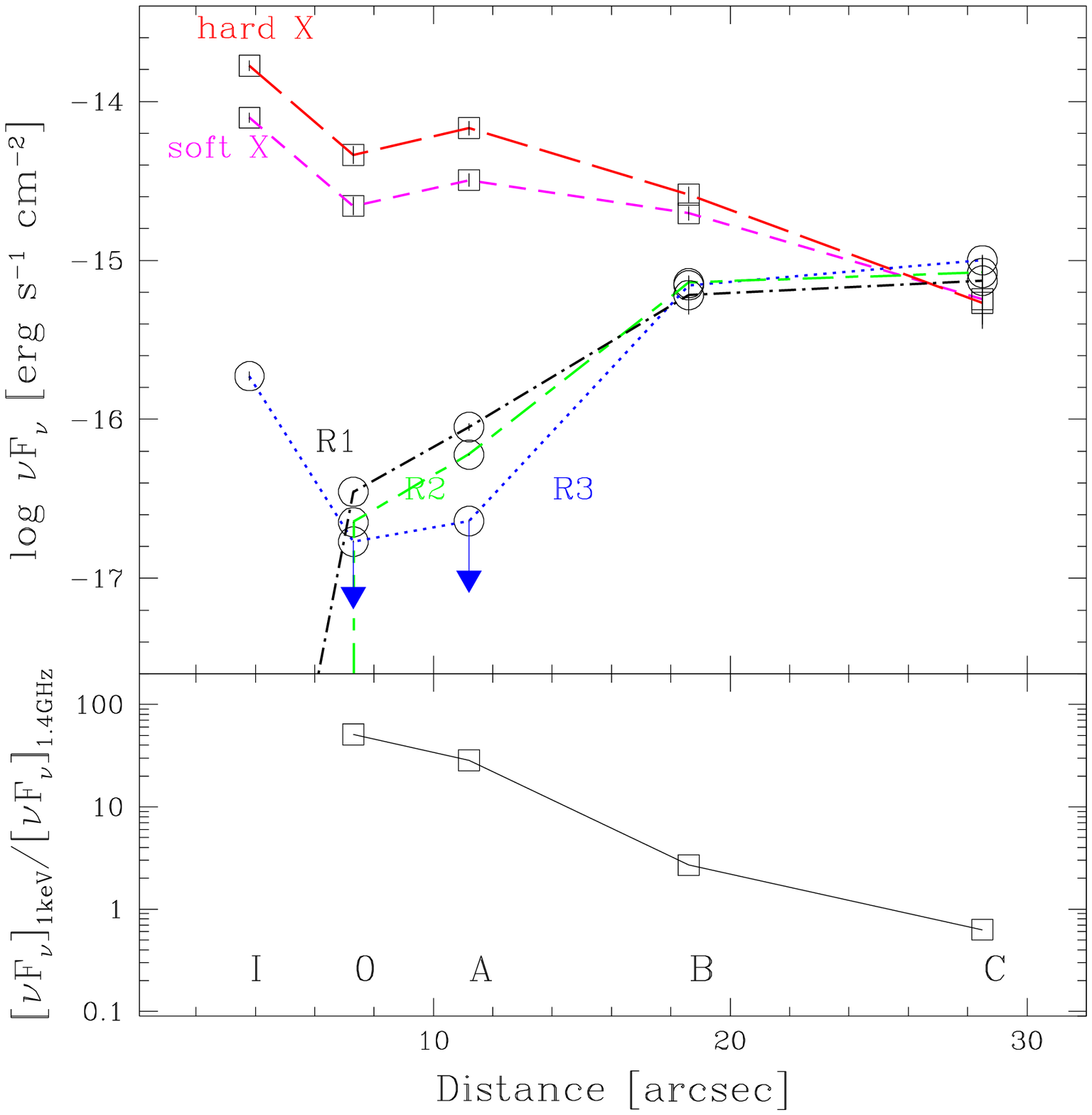}
\includegraphics[width=7cm]{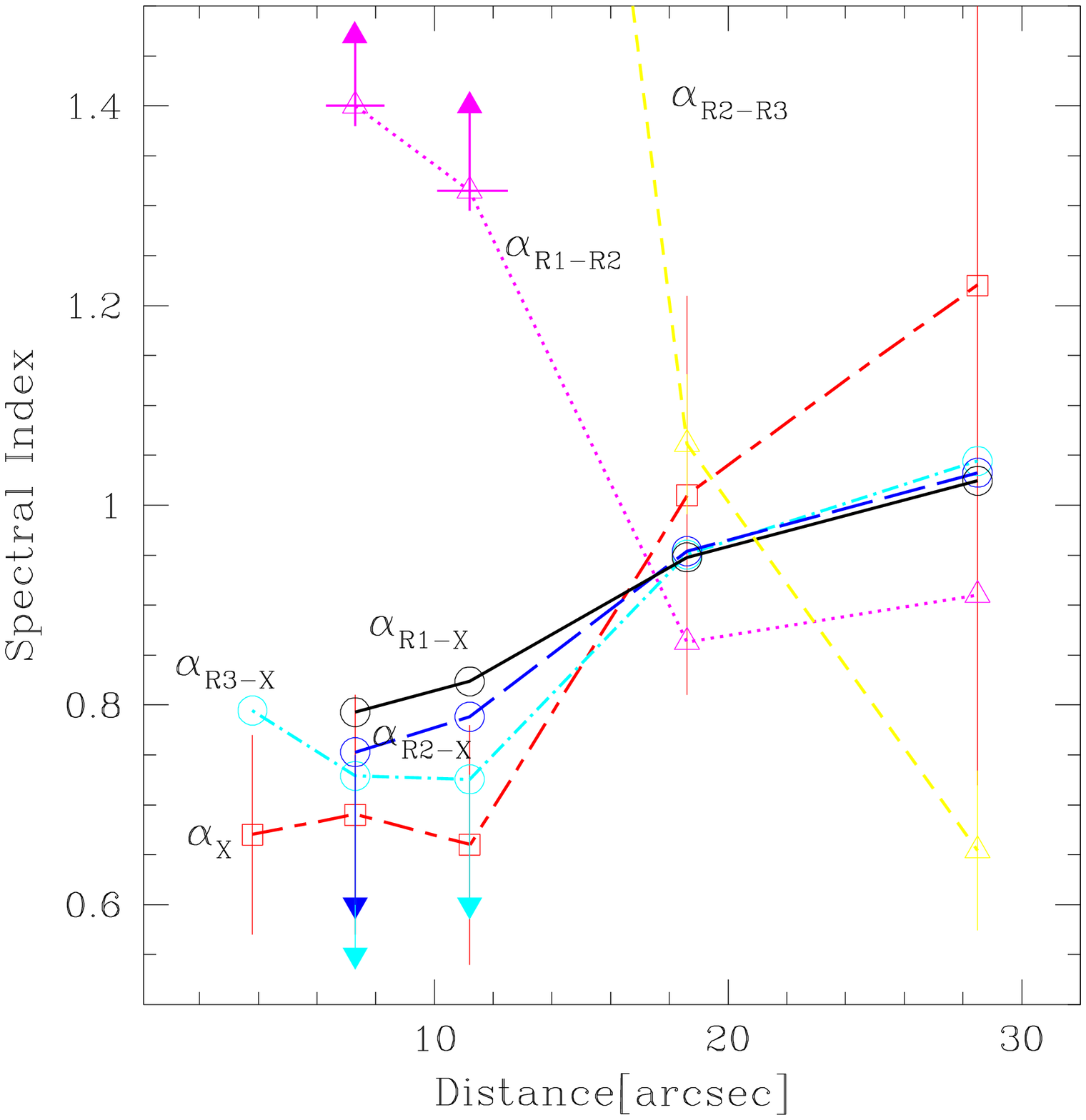}
\caption{{\bf Left:} Flux along the jet for each observed frequency: 
long dashed line -- hard X-rays (2-10~keV); short dashed line - soft
X-rays (0.5-2~keV); dotted line - R1 = 1.4 GHz; short-long-dashed - R2
= 5~GHz; and dash-dot - R3 = 8.5~GHz.  Lower panel shows the flux
ratio (1keV/1.4GHz) along the jet. {\bf Right:} Spectral index
($S_{\nu} \sim \nu^{-\alpha}$) along the jet: $\alpha_X$- X-ray
squares, and 
long-short dash; $\alpha_{R3-X}$ - 8.5~GHz to 1~keV
circles, 
dot-dashed line; $\alpha_{R2-X}$ - 5~GHz to 1~keV,
circles, 
dashed line; $\alpha_{R1-X}$ - 1.4~GHz to 1keV, circles,
solid line; $\alpha_{R1-R2}$ - 1.4~GHz to 5~GHz, triangles,
dotted line, $\alpha_{R2-R3}$ - 5~GHz to 8.5~GHz, triangles
dashed line.  }
\label{indices}
\end{figure}

\section{Jet Emission Models}

A detailed discussion of the model constraints can be found in
\citet{siemi2007}. Here we give a summary of the main results:

\subsection{One Zone Model Failures}

\subsubsection{Synchrotron} X-ray emission requires a continuous 
acceleration of high energy particles along the jet.  The expected
distribution of the freshly accelerated electrons gives $\alpha_X=
0.5$ as observed in the inner jet, but the radio spectrum is steep
with $\alpha_R > 1.0$ (Fig.3). A different cooling mechanism or
difference in electron populations emitting radio and X-rays is
needed.

\subsubsection{IC/CMB} models in the simplest version involve
an adiabatically expanding jet. In this case the observed
inverse-Compton (X-rays) and synchrotron (radio) intensity ratio is
different than the predicted one.  In the case of an efficient jet
confinement with some deceleration and B-field amplification it is
possible to get an agreement between the predicted and the observed
luminosity ratios, but the required B-field amplification is high
[$B(r_C)/B(r_o)$]$^2 \sim 50$. In this case for the equipartition at
$r_o$ there will be no equipartition at $r_C$.  The expected radio
spectrum should steepen more significantly than the X-ray
spectrum. The observed behavior is different: $\alpha_R$ decreases,
while $\alpha_X$ increases (Fig.3[right]).

\subsection{Two-Component Model: The Jet and a Sheath}

Radio and X-ray emissions are produced in two separate regions. The
X-ray emission comes from the proper relativistically moving jet and
the radio emission from the sheath - a slow moving radial extension of
the jet boundary layer.  The X-ray emission is due to synchrotron or
IC/CMB process.

\subsubsection{Proper Jet:}

The observed smooth decrease of the X-ray flux, when unconnected with
the increase of the radio one, is expected as a result of the decrease
in the inverse-Compton or synchrotron luminosities along decelerating
and/or expanding jet.

\subsubsection{Sheath:}

The sheath dominates the jet's radio emission and has two cooling
regions: (1) radiative cooling at the inner sheath: $t_{rad}<10$~Myr$<
t_{dyn}= 30 R_{10}$~Myr.  Due to the frequency dependent losses
a steep radio spectrum is expected in this regime.  (2) Adiabatic
cooling in the outer sheath: $t_{rad} > t_{dyn} < 10R_{10}$Myr. Because
of the frequency independent losses a flat spectrum is expected.

\subsection{Modulated Jet Activity?}

Can the morphology of the $>$300~kpc long jet be a result of the
intermittent activity of the quasar?  The knots are potentially too
large ($>$10~kpc) to be considered a result of extended shock waves
formed within a continuous jet outflow. They might form during
episodes of separate continuous activity. Then the size gives a
duration for each epoch of $\sim 10^5$ years.  The radio core can be
related to blazar phenomena \citep{blazej2004}, but the core also
resembles a compact GPS object at this time.

\section{Summary}   

The key results of this project:

\begin{itemize}

\item The jet is long: radio and X-ray jet emission is detected up to $\sim$300~kpc 
(projected distance) from the core.

\item The jet X-ray brightness is decreasing while the radio
brightness is increasing with distance from the quasar.

\item The one component X-ray/radio emission models failed to explain the
observations.

\item X-ray and radio jet properties suggests two separate
components for the jet emission.

\item Intermittent quasar activity might be
reflected in the observed jet morphology: each knot represents a
continuous jet activity with a duration of 10$^5$years.

\end{itemize}

\acknowledgements 

This research is funded in part by NASA contract NAS8-39073 and
Chandra Awards GO5-6113X, GO4-5131X, GO5-6118X, GO5-6111X. The
National Radio Astronomy Observatory is operated by Associated
Universities, Inc. under an agreement with the National Science
Foundation.









\end{document}